\documentclass[aps,prc,twocolumn,showpacs,showkeys,preprintnumbers,amsmath,amssymb,a4paper]{revtex4-1}

\usepackage{amssymb}
\usepackage{amsmath, bbm}
\usepackage[figuresright]{rotating}

\usepackage[figuresright]{rotating}

\newcommand{\MMBB}{
\!\begin{picture}(24,5)(0,0)
\put(6,1){\footnotesize $M_1M_2$}
\put(0,4){\line(3,-2){25}}
\put(0,-14){\footnotesize $B_{il}B_{ir}$}
\end{picture}
}
\newcommand{\MMB}{
\!\begin{picture}(24,5)(0,0)
\put(6,0){\footnotesize $M_1M_2$}
\put(0,3){\line(3,-2){25}}
\put(2,-13){\footnotesize $B_{i}$}
\end{picture}
}

\newcommand{\ccc}{{\cdot\cdot\cdot}}
\newcommand{\Kb}{{\overline{K}}}
\newcommand{\countzero}{\setcounter{equation}{0}%
         \setcounter{figure}{0}%
         \setcounter{table}{0}}

\def\La{\Lambda}
\def\Si{\Sigma}

\begin{document}

\title{Do $\Xi\Xi$ bound states exist?}
\author{J. Haidenbauer$^1$}
\author{Ulf-G. Mei{\ss}ner$^{2,1}$}
\author{S. Petschauer$^{3}$}
\affiliation{$^1$Institute for Advanced Simulation and J\"ulich
Center for Hadron Physics, Institut f\"ur
Kernphysik, Forschungszentrum J\"ulich, D-52425 J\"ulich, Germany\\
$^2$Helmholtz-Institut f\"ur Strahlen- und Kernphysik and Bethe
Center for Theoretical Physics, Universit\"at Bonn, D-53115 Bonn,
Germany \\
$^{3}$Physik Department, Technische Universit\"at M\"unchen, D-85747
  Garching, Germany
}

\begin{abstract}
The existence of baryon-baryon bound states in the strangeness sector is examined 
in the framework of SU(3) chiral effective field theory. Specifically, the role of SU(3) 
symmetry breaking contact terms that arise at next-to-leading order in the employed 
Weinberg power counting scheme is explored. 
We focus on the $^1S_0$ partial wave and on baryon-baryon channels with maximal 
isospin since in this case there are only two independent SU(3) symmetry
breaking contact terms. At the same time, those are the channels where
most of the bound states have been predicted in the past. 
Utilizing $pp$ phase shifts and $\Sigma^+ p$ cross section data allows us to pin 
down one of the SU(3) symmetry breaking contact terms and a clear indication for the 
decrease of attraction when going from the $NN$ system to strangeness $S=-2$ 
is found, which rules out a bound state for $\Si\Si$ with isospin $I=2$. 
Assuming that the trend observed for $S=0$ to $S=-2$ is not reversed when
going to $\Xi\Sigma$ and $\Xi\Xi$ makes also bound states in those systems rather 
unlikely.
\end{abstract}

\maketitle 

\section{Introduction}

Dibaryons (as compact six-quark systems or as bound states
formed by two conventional octet and/or decuplet baryons) have 
been intriguing objects of investigations and speculations for
many years. 
While in the purely nucleonic case there is yet again a promising 
dibaryon candidate \cite{Adlarson}, besides the deuteron, 
there are indications that the strangeness sector could be specifically 
rewarding for finding dibaryons \cite{Gal10}.
Here the by far best-known example is certainly the $H$-dibaryon suggested
by Jaffe \cite{Jaffe}, a deeply bound state with quantum numbers of the
$\Lambda\Lambda$ system, i.e. strangeness $S=-2$ and isospin $I=0$,
and with $J^P=0^+$. There are also speculations about the
existence of other exotic states, notably in the $S=-3$ ($N\Omega$) 
\cite{Goldman3,Etminan} and $S=-6$ ($\Omega\Omega$) \cite{Zhang,Goldman6}
systems, see also Refs.~\cite{Buchoff2012,Yamada2014}.

With regard to two octet baryons, the (approximate) SU(3) flavor symmetry of 
the strong interaction suggests that bound states could exist for systems
with strangeness $S=-3$ and, in particular, $S=-4$ \cite{Miller}.
Indeed, meson-exchange models like the Nijmegen baryon-baryon ($BB$)
interaction \cite{Rij99,Rij10}, derived under the assumption of 
(broken) SU(3) symmetry, predict interactions for the $S=-3$ and $-4$ sectors 
that are fairly strong and attractive and lead to bound states in 
the $\Xi\Sigma$ and $\Xi\Xi$ channels \cite{Sto99,Rij13}.
The situation is somewhat different for $BB$ interactions derived in the 
constituent quark model by Fujiwara and collaborators \cite{fss2}. 
While SU(3) flavor symmetry plays likewise a key role in extending the model 
from the $NN$ and $YN$ interaction (where free parameters are fixed) to 
the $S=-3$ and $-4$ channels, in this approach it was found that the 
$BB$ interaction becomes step by step less attractive when going from 
strangeness $S=0$ to $S=-4$.
In particular, no dibaryon bound states are supported, except for the deuteron.
A similar pattern was reported in Ref.~\cite{Reu96} where the intermediate-range 
attraction from the scalar-isoscalar (``$\sigma$'') channel was evaluated within a 
model for correlated $\pi\pi$ and $\bar KK$ exchange between octet baryons. Also 
in this case it was found that the attraction between two baryons, quantified by 
the effective $\sigma$-meson coupling strength, decreases step by step in the
strangeness sector.  

Results obtained in lattice QCD calculations are conflicting so far. While 
a $\Xi\Xi$ bound state was found by the NPLQCD collaboration~\cite{Beane} (in 
the $^1S_0$ state), the HAL QCD collaboration reported only a moderately 
attractive interaction for that partial wave \cite{Sasaki}.

In the present paper we examine the existence of $\Si\Si$, $\Xi\Si$ and $\Xi\Xi$ bound 
states in the framework of SU(3) chiral effective field theory (EFT). In particular, we 
explore the role of SU(3) symmetry breaking contact terms that arise at next-to-leading 
order (NLO) in the perturbative expansion of the baryon-baryon potential. 
A first study of the baryon-baryon ($BB$) interactions within chiral EFT \cite{Po06} 
in the Weinberg scheme \cite{Wei90,Wei91}
for the strangeness $S=-2$, $-3$ and $-4$ sectors was presented in Refs.~\cite{Po07,Hai10}. 
At leading-order (LO) considered in those works the chiral potentials consist of contact
terms without derivatives and of one-pseudoscalar-meson exchanges ($\pi$, $K$, $\eta$).
Assuming ${\rm SU(3)}$ flavor symmetry those contact terms and the couplings of the
pseudoscalar mesons to the baryons can be related to the corresponding quantities 
of the $S=-1$ hyperon-nucleon ($YN$) channels. Specifically, the values of the 
pertinent five low-energy constants (LECs) related to the contact terms could be fixed from
the study of the $\Lambda N$ and $\Sigma N$ systems \cite{Po06} and then 
genuine predictions for the $\Xi\Lambda$, $\Xi\Sigma$, and $\Xi\Xi$ interactions 
could be made at LO. Strong attraction was found in some of the $S=-2$, $-3$ and $-4$ 
$BB$ channels, and several bound states were predicted \cite{Hai10}.

Recently, a $YN$ interaction has been derived up to NLO in chiral EFT by 
the J\"ulich-Bonn-Munich group \cite{Hai13}. At that order contact terms leading to 
an explicit SU(3) symmetry breaking appear for the first time \cite{Hai13,Pet13} as
mentioned above. 
Since the sparse experimental information on $\Lambda N$ and $\Sigma N$ scattering could be 
described rather well with using the SU(3) symmetric terms alone,
SU(3) symmetry breaking was simply neglected. In other words it was assumed that
the LECs associated with those contact terms are zero. 
Thus, in the actual calculation the SU(3) symmetry is only broken via the employed
physical masses of the involved mesons and octet baryons 
($N$, $\Lambda$, $\Sigma$, $\Xi$). 

On the other hand, it was also found in Ref.~\cite{Hai13} that a simultaneous description
of the $YN$ data and the nucleon-nucleon ($NN$) phase shifts is not possible on
the basis of SU(3) symmetric contact terms. In particular, the strengths needed for
reproducing the $pp$ (or $np$) $^1S_0$ phase shifts and the $\Sigma^+ p$ cross section
could not be reconciled in a scenario which maintained SU(3) symmetry for the contact
terms. This observation is the starting point for the present study, because it can be
used to put constraints on the SU(3) symmetry breaking contact terms. In particular,
the situation in the $^1S_0$ partial wave and for $BB$ channels with maximal isospin
is rather simple and interesting. Here, there are only two independent SU(3) symmetry
breaking LECs at NLO for five physical channels, and for three of those five
channels bound states have been predicted in the past. 
The aforementioned $pp$ phase shifts and the $\Sigma^+ p$ cross section allow us to 
pin down one of the symmetry breaking LECs and provide a clear-cut indication for the 
decrease of attraction when one goes from the $NN$ system to $S=-2$, so that a bound 
state for $\Si\Si$ with isospin $I=2$ can be practically ruled out.
The other LEC cannot be determined
at present and several options for its value are discussed. However, already the
assumption that the trend one sees for $S=0$ to $S=-2$ is not reversed when
going to $S=-3$ and $S=-4$ makes bound states in the latter systems rather unlikely.

The paper is structured in the following way: 
In Sect. 2 we provide a basic introduction to our $BB$ interaction derived in chiral EFT. 
We also discuss the changes that arise in the interaction when the SU(3) symmetry
breaking contact terms are taken into account. 
Selected results for $BB$ systems with strangeness $S=-2$ to $S=-4$ based on SU(3)
symmetric contact terms fixed in a fit to $YN$ data are presented in Sect. 3. 
In Sect. 4 we introduce the SU(3) symmetry breaking contact terms and show the
implications for the $^1S_0$ phase shift in the $\Si\Si$, $\Xi\Si$, and 
$\Xi\Xi$ channels with maximal isospin.
The paper ends with a short summary. 
Some technical information about our calculation is given in Appendix A. 

\section{The baryon-baryon interaction in chiral EFT}

A comprehensive description of the derivation of the chiral $BB$ 
potentials for the strangeness sector using the Weinberg power counting can be
found in Refs.~\cite{Po06,Hai13,Pet13,Haidenbauer:2007ra}.
The LO potential consists of four-baryon contact terms without derivatives and of
one-pseudoscalar-meson exchanges while at NLO contact terms with two derivatives
arise, together with contributions from (irreducible) two-pseudoscalar-meson exchanges.
The interaction in Ref.~\cite{Hai13} was derived by imposing SU(3) flavor symmetry.
Then the contributions from pseudoscalar-meson exchanges ($\pi$, $\eta$, $K$) are 
completely fixed in terms of the axial coupling $g_A$ and $\alpha$, 
the so-called $F/(F+D)$ ratio, together with the pion decay constant $f_0$. 

SU(3) symmetry was also imposed for the contact terms. Since the strength parameters 
associated with the contact terms, the LECs, need to be determined by a fit to data,
it was tried to keep the number of independent LECs that can contribute as small
as possible. In the SU(3) symmetric case there are in total 13 LECs entering the 
$S$-waves and the $S$--$D$ transitions of the $\La N$--$\Si N$ system \cite{Hai13},
and their values could be fairly well fixed in a fit to the available low-energy 
total cross sections for $\Lambda p \to \Lambda p$, $\Sigma^-p \to 
\Lambda n$, $\Sigma^-p \to \Sigma^0 n$, $\Sigma^-p \to \Sigma^- p$,
and $\Sigma^+p \to \Sigma^+ p$.
However, it would have been not possible to determine the additional 5 contact terms 
appearing at NLO that lead to an explicit SU(3) symmetry breaking, cf. the Appendix 
of Ref.~\cite{Hai13} and also Ref.~\cite{Pet13}, and, therefore, the corresponding
LECs were simply set to zero. 
 
At the same time, it became already clear in Ref.~\cite{Hai13} that it is 
impossible to obtain a combined fit to the $YN$ data and to the $NN$ phase shifts
with LECs that fulfill SU(3) symmetry. The most obvious case is the $^1S_0$ partial 
wave, where SU(3) symmetry implies that the interactions in the $NN$ (I=1) 
and $\Sigma N$ (I=3/2) channels involve the very same two LECs and are given 
simply by 
\begin{equation}
V (^1S_0) = \tilde C^{27}_{^1S_0} + C^{27}_{^1S_0}(p^2+p'^2) \ , 
\label{LECSU3}
\end{equation}
with $p$ and $p'$ being the center-of-mass momenta in the initial and final state.
The label $\{27\}$ indicates that both channels belong to the $\{27\}$ 
representation of SU(3) \cite{Swa63,Dover91}, see Ref.~\cite{Hai13} for a 
detailed description of the notation. 
Indeed, strict SU(3) symmetry suggests that the $^1S_0$ contact interaction should 
be the same for several $BB$ channels that belong solely to the $\{27\}$: 
\begin{equation}
V^{(I=1)}_{NN} =  V^{(I=3/2)}_{\Sigma N} =
V^{(I=2)}_{\Sigma\Sigma} =
V^{(I=3/2)}_{\Xi\Sigma} = V^{(I=1)}_{\Xi\Xi} \ .
\label{LECSU3a}
\end{equation}

Eq.~(\ref{LECSU3}) implies that under the assumption of SU(3) symmetry 
the (hadronic part of the) interaction in the $\Sigma^- n$ or $\Sigma^+ p$ 
channels is unambiguously fixed once the LECs are 
determined from the $np$ or $pp$ phases, or vice versa. 
In practice it turned out that with LECs fixed from the $np$ (or $pp$) phase shifts a 
near-threshold bound state is generated in the $\Sigma^+ p$ system and, as a 
consequence, the empirical $\Sigma^+ p$ cross section is grossly 
overestimated \cite{Hai13}. Evidently, SU(3) symmetry breaking in the contact terms has to be
taken into account if one wants to describe $NN$ and $\Si N$ scattering simultaneously.

The contact terms, including the SU(3) symmetry breaking corrections that arise at NLO, 
have been worked out explicitly in Ref.~\cite{Pet13} for all octet $BB$ channels
from strangeness $S=0$ to $-4$. There are twelve independent SU(3) symmetry breaking LECs in total, 
see Ref.~\cite{Pet13}, of which six occur in the $^1S_0$ partial wave
and the other six in the $^3S_1$. It is impossible to determine all of those based on the
presently available experimental information in the strangeness $S=-1$ to $-4$ sectors.

The situation is more favorable, however, in the particular case discussed above, namely
for the $^1S_0$ partial wave and $BB$ channels with maximal isospin. Here one 
obtains
\begin{eqnarray}
\nonumber
V^{(I=1)}_{N N} &=& \tilde C^{27}_{^1S_0} + C^{27}_{^1S_0}(p^2+p'^2)
+\frac{1}{2}C_1^\chi (m^2_K-m^2_\pi) ,\\
\nonumber
V^{(I=3/2)}_{\Sigma N } &=& \tilde C^{27}_{^1S_0} + C^{27}_{^1S_0}(p^2+p'^2)
+\frac{1}{4}C_1^\chi (m^2_K-m^2_\pi) ,\\
\nonumber
V^{(I=2)}_{\Sigma \Sigma } &=& \tilde C^{27}_{^1S_0} + C^{27}_{^1S_0}(p^2+p'^2) , \\
\nonumber
V^{(I=3/2)}_{\Xi \Sigma } &=& \tilde C^{27}_{^1S_0} + C^{27}_{^1S_0}(p^2+p'^2)
+\frac{1}{4}C_2^\chi (m^2_K-m^2_\pi) , \\
\nonumber
V^{(I=1)}_{\Xi\Xi } &=& \tilde C^{27}_{^1S_0} + C^{27}_{^1S_0}(p^2+p'^2)
+\frac{1}{2}C_2^\chi (m^2_K-m^2_\pi) . \\ 
\label{LEC}
\end{eqnarray}
Evidently, there are only two additional LECs due to SU(3) symmetry breaking, which are 
denoted by $C^\chi_1$ and $C^\chi_2$ in the above equations. As expected, their contributions 
are proportional to the meson mass difference, $m^2_K-m^2_\pi$, so that they vanish in case of
SU(3) symmetry, i.e. when $m^2_K \equiv m^2_\pi$.
There are also contact terms proportional to $m_\pi^2$ and $m_K^2$ which are, however,
SU(3) symmetric and have been absorbed into the definition of $\tilde C^{27}_{^1S_0}$ \cite{Pet13}.
A combined fit to the $pp$ (or $np$) $^1S_0$ phase shifts and the $\Si^+ p$
cross section allows us to determine three of the four LECs in Eq.~(\ref{LEC}), 
as will be demonstrated below. Then we can make genuine predictions for the $\Si\Si$ interaction 
with isospin $I=2$. The fourth LEC ($C^\chi_2$) cannot be pinned down reliably at present. In this 
case we will consider a range of values and study the consequences for the possible existence 
of $\Xi\Si$ and $\Xi\Xi$ bound states. 

For completeness let us mention that our calculations are done in momentum space. We solve 
the partial-wave projected (non--relativistic) Lippmann-Schwinger (LS) equation,
\begin{widetext}
\begin{equation}
T_{B_1B_2}\,(p'',p';\sqrt{s})=V_{B_1B_2}\,(p'',p')+
\int_0^\infty \frac{dpp^2}{(2\pi)^3} \, V_{B_1B_2}\,(p'',p)
\frac{2\mu_{B_1B_2}}{k^2-p^2+i\epsilon}T_{B_1B_2}\,(p,p';\sqrt{s})\ 
\label{LSE} 
\end{equation}
\end{widetext}
for a particular $BB$ channel. Here, $\mu_{B_1B_2}$ is the reduced mass and $k$ is
the on-shell momentum, which is defined by
$\sqrt{s}=\sqrt{M^2_{B_{1}}+k^2}+\sqrt{M^2_{B_{2}}+k^2}$.
Relativistic kinematics is used for relating the laboratory momentum $p_{{\rm lab}}$
of the baryons to the center-of-mass momentum. In case of $pp$ and $\Sigma^+p$,
where we compare with experiments, the Coulomb interaction is included. This is 
done via the Vincent-Phatak method \cite{VP}. 
The potentials in the LS equation are cut off with a regulator function, $f_R(\Lambda) =
\exp\left[-\left(p'^4+p^4\right)/\Lambda^4\right]$,
in order to remove high-energy components \cite{Epe05}.
In Ref.~\cite{Hai13} results for cutoff values in the range $\Lambda=500$ -- $650\,$MeV 
were shown and we will consider the same range here. The variation of the results with 
the cutoff can be viewed as a rough estimate for the theoretical uncertainty \cite{Epe05}.
A better method to determine the theoretical uncertainty has recently been proposed
for the $NN$ sector \cite{Epelbaum:2014efa},  but in view of the scarce data in the strangeness sector and given the
exploratory character of our study, we stick to the much simpler procedure of varying the cutoff.
 
Note that for all the systems listed in Eq.~(\ref{LEC}) there is no coupling to other
partial waves or channels. Thus, differences in the reaction thresholds that generate
an additional SU(3) symmetry breaking in the scattering amplitude when the
LS equation (\ref{LSE}) is solved for coupled channels, are absent.
This makes those systems especially suited for isolating SU(3) symmetry breaking
effects in the potential. 
 
\begin{table*}
\caption{
$\Sigma\Sigma$, $\Xi\Sigma$ and $\Xi\Xi$ scattering lengths (in fm) in the $^1S_0$ partial 
wave. Results are given for our LO \cite{Po06} and NLO \cite{Hai13} interactions based 
on LECs fitted to the $YN$ data. For comparison some values for the Nijmegen 
NSC97 potential \cite{Sto99} and a quark model \cite{fss2} are also included. 
}
\vskip 0.2cm 
\centering
\renewcommand{\arraystretch}{1.3}
\begin{tabular}{|c|rr|rr|r|}
\hline
& {$\chi$EFT LO} & {$\chi$EFT NLO}& {NSC97a \cite{Sto99}} & {NSC97f \cite{Sto99}} & {fss2 \cite{fss2}} \\
\hline
${\Lambda}$ [MeV]& 550$\ccc$700 &500$\ccc$650 &  &  &  \\
\hline
$a^{I=2}_{\Sigma\Sigma}$ &$-6.2\ccc -9.3$   &$60.6\ccc -286.0$  & $10.32$ & $6.98$ & $-85.3$ \\
\hline
\hline
$a_{\Xi\Lambda}$ &$-33.5\ccc 9.07$     &$-7.4\ccc -13.5$   & $-0.80$ & $-2.11$ & $-1.08$\\
\hline
$a^{I=3/2}_{\Xi\Sigma}$  &$4.28\ccc 2.74$ &$ 8.4\ccc 13.8$   & $4.13$ & $2.32$ & $-4.63$\\
\hline
\hline
$a^{I=1}_{\Xi\Xi}$ &$3.92\ccc 2.47$     &$ 9.7\ccc  6.5$   & $17.81$ & $2.38$ & $-1.43$\\
\hline
\end{tabular}
\renewcommand{\arraystretch}{1.0}
\label{tab:T1}
\end{table*}

\section{Results based on the low energy constants of our NLO $YN$ potential}

In this section we present predictions for the $\Sigma\Sigma$, $\Xi\Sigma$ 
and $\Xi\Xi$ channels, where SU(3) symmetry is assumed for the contact terms. 
To be exact, SU(3) symmetry is utilized to relate the LECs for the $S=-2$, $-3$ and $-4$
systems to those determined in the fit to the $\La N$ and $\Si N$ data \cite{Hai13}. 
The symmetry is broken by the used physical masses of the involved mesons and 
baryons. Note that the meson masses induce an explicit symmetry breaking into 
the $BB$ potential while the baryon masses enter only in the course of solving the
scattering equation, because they appear in the integral equation 
in form of the reduced mass, see Eq.~(\ref{LSE}). 
 
In a corresponding investigation with our LO potential it was found that the 
interaction in some of the $S= -3$ and $-4$ channels is strongly attractive and 
even bound states were predicted \cite{Po07,Hai10}. 
The same happens also at NLO as one can see from the results for the $^1S_0$
partial wave summarized in Tables~\ref{tab:T1} and \ref{tab:T2}. Specifically, in all 
channels where large scattering lengths were found at LO, they are likewise
large at NLO. And, except for $\Xi \Lambda$, the scattering lengths are large and
positive -- a clear indication for bound states. In case of $\Sigma\Sigma$ the NLO
interaction produces a pole very close to the threshold which, depending on the cutoff,
corresponds either to a bound state (large positive scattering length) or to a virtual
state (large negative scattering length). The actual binding energies of
those states are listed in Table~\ref{tab:T2}. Comparing the NLO results with the
ones at LO one notices that the $\Xi\Xi$ and $\Xi\Sigma$ binding energies have become 
somewhat smaller. Indeed in both cases the systems are now only fairly weakly bound.
We do not include the Coulomb interaction in the calculation of the
strangeness $S= -2$ to $-4$ sectors. (For bound states this would be 
technically rather complicated within the Vincent-Phatak method employed by us.)
It is quite possible that the additional repulsion due to the Coulomb force could 
even make the $\Si\Si$ system unbound.
In this context we want to point out that it is good to see that the cutoff 
dependence of the binding energies is strongly reduced at NLO. 

Table~\ref{tab:T1} contains also the scattering lengths predicted by the 
Nijmegen NSC97 meson-exchange model \cite{Rij99} and of a $BB$ potential by
Fujiwara and collaborators \cite{fss2} derived in the quark model. The Nijmegen
interaction suggests bound states in the $\Sigma\Sigma$, $\Xi\Sigma$ and $\Xi\Xi$
channels, as can be guessed from the large and positive scattering length. 
The latest version of the Nijmegen potential \cite{Rij10} produces a bound state 
in the $\Xi\Xi$ channel \cite{Rij13} too. 
As already mentioned in the Introduction, no bound states were found for the
quark-model interaction \cite{fss2}, though the $\Sigma\Sigma$ interaction is 
seemingly very close to producing a bound state as indicated by the large
negative scattering length. 

\begin{table}[h]
\centering
\caption{
Binding energies of various $BB$ bound states (in MeV) in the $^1S_0$ partial wave,
for our LO \cite{Po06} and NLO \cite{Hai13} interactions based 
on LECs fitted to the $YN$ data.}
\vskip 0.2cm 
\renewcommand{\arraystretch}{1.3}
\begin{tabular}{|c|c|c|}
\hline
& {$\chi$EFT {LO}} &  {$\chi$EFT {NLO}} \\
\hline
${\Lambda}$ [MeV] & 550 {$\ccc$} 700 & 500 {$\ccc$} 650 \\
\hline
$\Sigma\Sigma$\, $(I=2)$ & $ -  $ & $ 0$ {$\ccc$} $-0.01$ \\ 
$\Xi\Sigma$\, $(I=3/2)$ & $-2.23$ {$\ccc$} $-6.18$  & $-0.58$ {$\ccc$} $-0.19$ \\
$\Xi\Xi$\, $(I=1)$    & $-2.56$ {$\ccc$} $-7.27$ & $-0.40$ {$\ccc$} $-1.00$ \\
\hline
\end{tabular}
\renewcommand{\arraystretch}{1.0}
\label{tab:T2}
\end{table}

\section{Results with inclusion of SU(3) symmetry breaking contact terms}

For studying the effects of SU(3) symmetry breaking we performed fits to $NN$ and $YN$ 
data, requiring that $C^{27}$ is the same in line with the power counting where 
(SU(3) symmetry breaking) corrections to $\tilde C^{27}$ arise at NLO but not to $C^{27}$. 
With regard to $NN$ the fit was performed to the $^1S_0$ $pp$ phase shifts of the GWU 
analysis \cite{SAID,SAID1}.
Since the $pp$ interaction is slightly less attractive than the one in $np$, cf. the
scattering lengths of $\approx$ $-17$ fm (for the purely hadronic $pp$ interaction) 
versus $-23.75$ fm, the amount of SU(3) symmetry breaking we need to introduce 
is also somewhat smaller.
The LEC $C^{27}$ was determined in the $pp$ sector and then taken over in the subsequent
calculations in the strangeness sector. It turned out that the actual value of $C^{27}$
found in the fits depends only very weakly on the cutoff mass and, therefore, we 
adopted a single value for all cutoffs. 

The $\Sigma^+ p$ interaction was fitted to the corresponding $^1S_0$ phase shift 
predicted by our chiral EFT $YN$ interaction \cite{Hai13}. We could not simply take 
over the results of Ref.~\cite{Hai13} because that interaction is based on a single
decay constant $f_0 \approx f_\pi \approx$ 93 MeV. Now we want to take into account also
the experimentally known differences between $f_\pi$, $f_\eta$, and $f_K$ in the
evaluation of the pertinent coupling constants. 
In the fit we made sure that there is perfect agreement with the results of \cite{Hai13}
in the (low-energy) region where $\Sigma^+ p$ cross section data are available.
In fact, for one cutoff ($\Lambda =$ 600 MeV) we even performed a full fit to all
$YN$ data considered in \cite{Hai13} in order to check whether the same $\chi^2$ can 
be achieved -- which was indeed the case.

\begin{figure}[t]
\begin{center}
\includegraphics[height=8.cm,angle=-90,keepaspectratio]{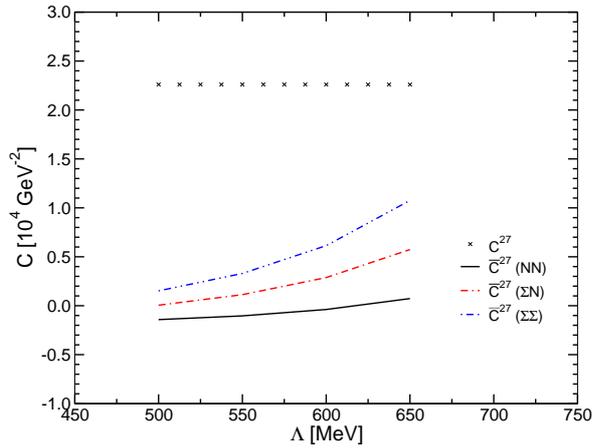}
\end{center}
\caption{Low-energy constants employed in the different $BB$ channels, 
for the considered cutoff values $\Lambda$.
Here ${\overline C}^{27}(NN)=\tilde C^{27}+\frac{1}{2}C^\chi_1 (m^2_K-m^2_\pi)$, etc.,
see Eq.~(\ref{LEC}).
$C^{27}$ is in units of 10$^4$ GeV$^{-4}$. 
}
\label{fig:LEC}
\end{figure}

\begin{figure*}
\begin{center}
\includegraphics[height=8.0cm,keepaspectratio]{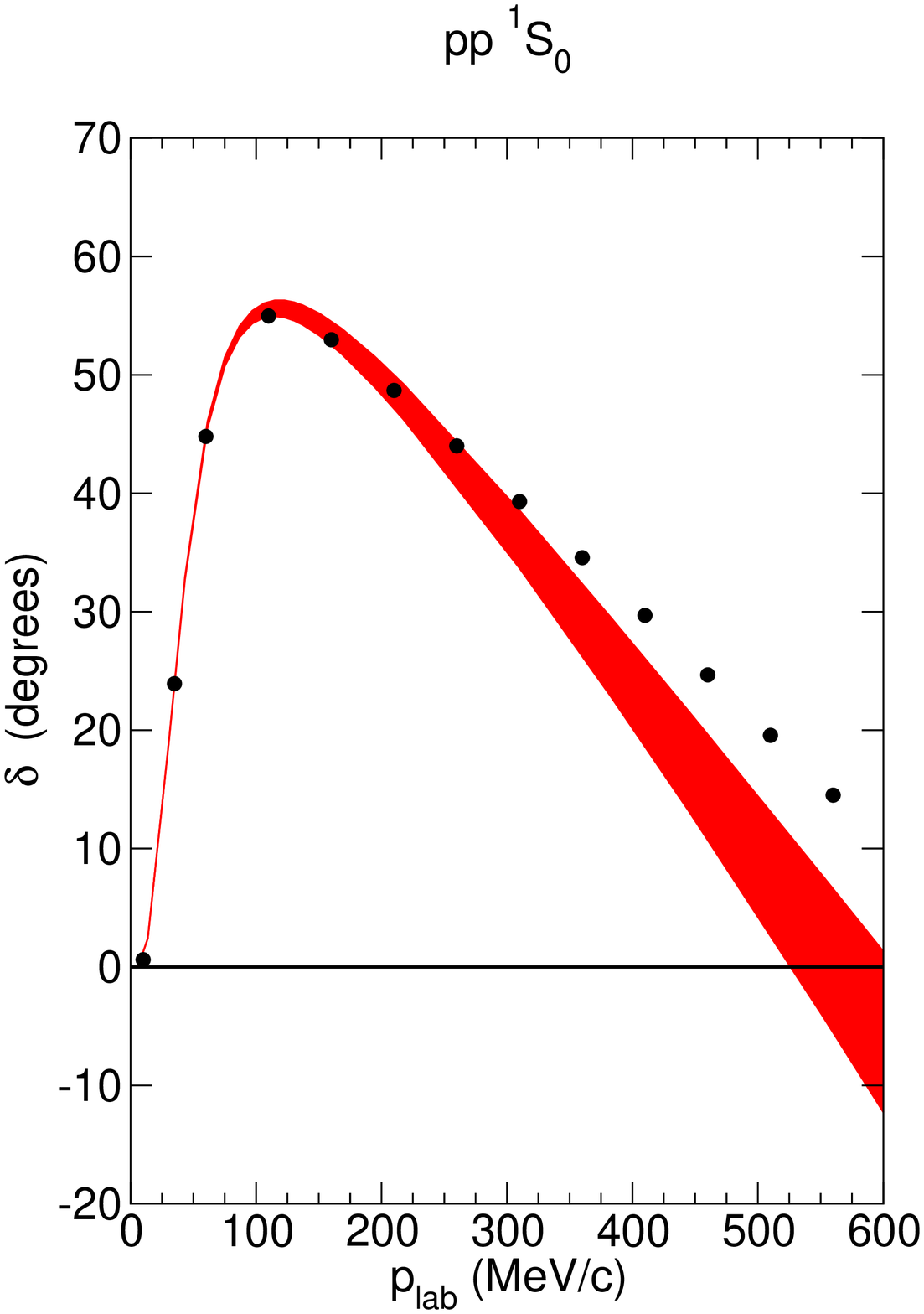}\includegraphics[height=8.0cm,keepaspectratio]{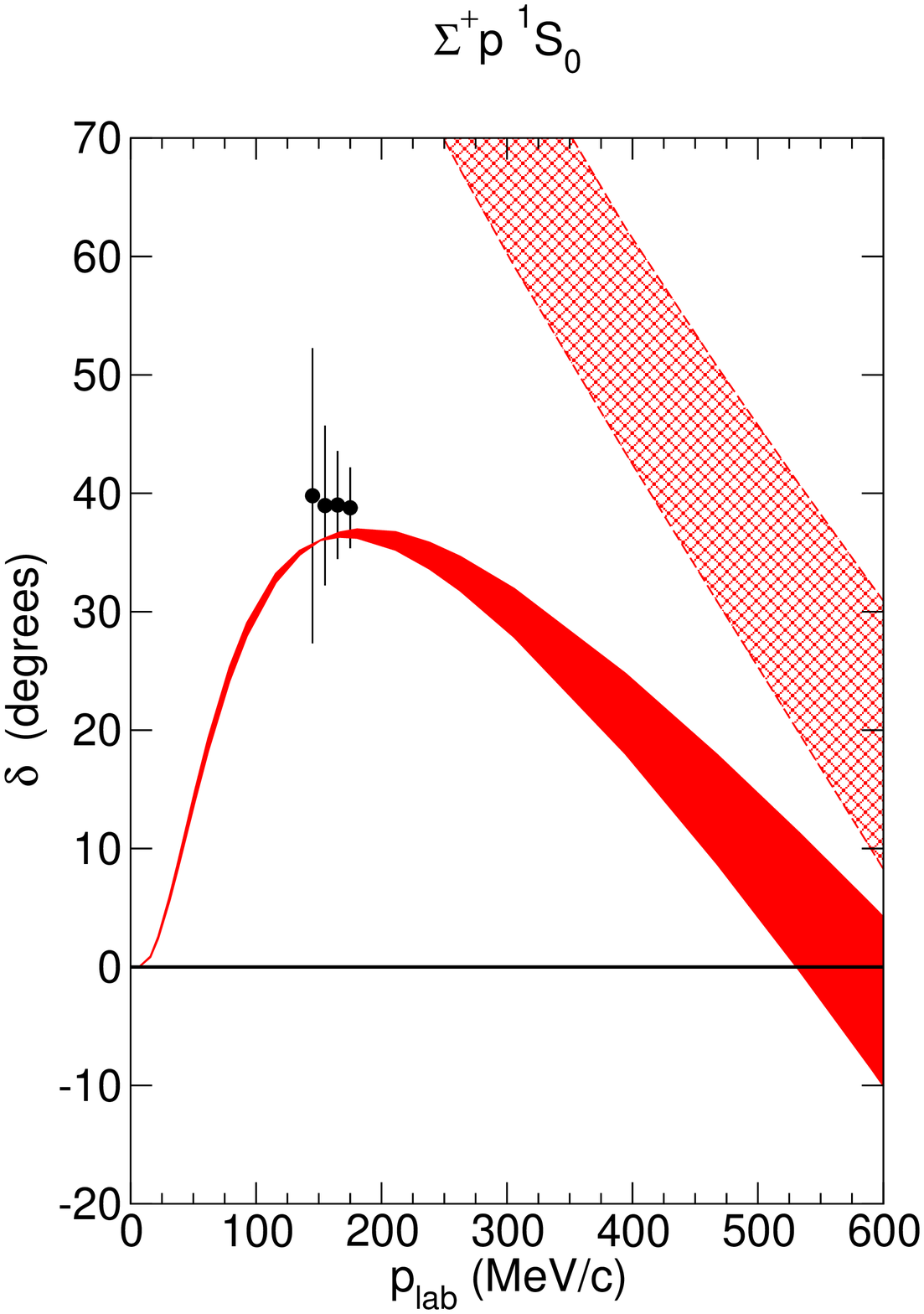}
\end{center}
\caption{$pp$ and $\Sigma^+p$ phase shifts in the $^1S_0$ partial wave. The filled band represent our 
results at NLO. The hatched band shows $\Sigma^+p$ result based on LECs fixed by a fit to $pp$ 
phase shifts. The $pp$ phase shifts of the GWU analysis \cite{SAID1} are shown by circles.
In case of $\Sigma^+ p$ the circles indicate upper limits for the phase shifts, 
deduced from the $\Sigma^+ p$ cross section, see text. 
}
\label{fig:NN}
\end{figure*}

It turns out that $C^{\chi}_1 < 0$, i.e. one needs more repulsion to fit $\Sigma^+p$ ($Y N$) 
data than to fit the $pp$ $^1S_0$ phase shift. The LECs are graphically presented in 
Fig.~\ref{fig:LEC}, while the phase shifts are shown in Fig.~\ref{fig:NN}.
For the former we show the sum of the LO contact term
${\tilde C}^{27}$ and the SU(3) symmetry breaking contribution for each $BB$
channel, e.g. ${\overline C}^{27}=\tilde C^{27}+\frac{1}{2}C^\chi_1 (m^2_K-m^2_\pi)$ for
the $NN$ system, so that one can see how the repulsion effectively increases when
going from $S=0$ to $-2$. 
The values of the employed LECs are summarized in Table~\ref{tab:LECs}. 
 
\begin{table}[h]
\centering
\caption{Employed low energy constants for various cutoffs. The values 
for $\tilde C^{27}$ are in 10$^4$ GeV$^{-2}$, those for 
$C^{27}$ and $C^\chi_1$ in 10$^4$ GeV$^{-4}$.
}
\vskip 0.2cm 
\renewcommand{\arraystretch}{1.3}
\begin{tabular}{|c|ccc|}
\hline
$\Lambda$ (MeV) & {$\tilde C^{27}$} & {$C^{27}$} &  {$C^\chi_1$} \\
\hline
500 & 0.15196 & 2.26 &  -2.6014    \\
550 & 0.32963 & 2.26 &  -3.8346    \\
600 & 0.61394 & 2.26 &  -5.7731   \\
650 & 1.0752  & 2.26 &  -8.8719   \\
\hline
\end{tabular}
\label{tab:LECs} 
\renewcommand{\arraystretch}{1.0}
\end{table}

The experimental $\Sigma^+ p$ cross section provides an upper limit on the phase shift 
for the $\Sigma^+ p$ $^1S_0$ partial wave. The limit can be derived from the 
expression for the partial cross section, 
\begin{equation}
\sigma_{\Sigma^+ p;\, J} = \frac{(2J+1)\pi}{k^{\,2}}  \, \sin^2 \delta_J, 
\label{Partial}
\end{equation}
$J$ being the total angular momentum, by assuming that the $^1S_0$ contribution 
alone already saturates the cross section data. Pertinent results are included in 
Fig.~\ref{fig:NN}, see the filled circles. Obviously for our EFT interaction 
\cite{Hai13} (but also for most of the $YN$ potentials based on meson 
exchange \cite{Rij99,Rij10,Hol89,Hai05}) 
the predicted $^1S_0$ amplitude is very close to saturating the $\Sigma^+ p$ cross 
section alone. The hatched band in Fig.~\ref{fig:NN} indicates the predictions
one would get for the $\Sigma^+ p$ channel with the LECs fitted to 
the $pp$ $^1S_0$ phase shifts. Evidently, the assumption of SU(3) symmetry for 
the contact terms is in clear contradiction with the experimental information.

Note that there is also a phase shift analysis for $\Sigma^+ p$ \cite{Nagata} 
at a single momentum, namely $p_{lab} = 170$ MeV/c, which suggests a value of 
around 26 degrees for the $^1S_0$ partial wave. However, that analysis is not 
model independent and, therefore, we have more confidence in our own results 
determined by a fit to existing $YN$ data within chiral EFT. 

\begin{figure}
\begin{center}
\includegraphics[height=8.0cm,keepaspectratio]{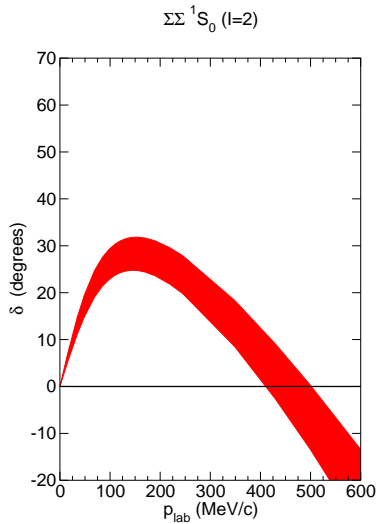}
\end{center}
\caption{Phase shifts for the $^1S_0$ partial wave in the $\Sigma\Sigma$ channel with isospin $I=2$. 
The band is our prediction based on the LECs $\tilde C^{27}$ and $C^{27}$ fixed from a fit 
to $pp$ and $\Sigma^+ p$, see Eq.~(\ref{LEC}). 
}
\label{fig:SS}
\end{figure}

Once we have determined $\tilde C^{27}$, $C^{27}$, and $C^\chi_1$ from our fit to 
the $pp$ and $\Sigma^+ p$ $^1S_0$ phase shifts, we can make predictions for 
the $\Sigma\Sigma$ case, see Eq.~(\ref{LEC}). Corresponding results are shown in 
Fig.~\ref{fig:SS}. The phase shifts attest that there is a sizable attraction in
this channel but the actual values are in the order of 30 degrees and, thus,
far away from the SU(3) symmetric case discussed in Sect. 2 where the 
$\Sigma\Sigma$ system with $I=2$ was more or less bound. In particular, the
predicted scattering lengths are now around $-3.2$ to $-3.4$ fm only. 
Indeed, the present result at NLO that follows directly from the SU(3) symmetry 
breaking observed between $pp$ and $\Sigma^+ p$ practically rules out a bound 
state in this channel.  

\begin{figure*}
\begin{center}
\includegraphics[height=8.0cm,keepaspectratio]{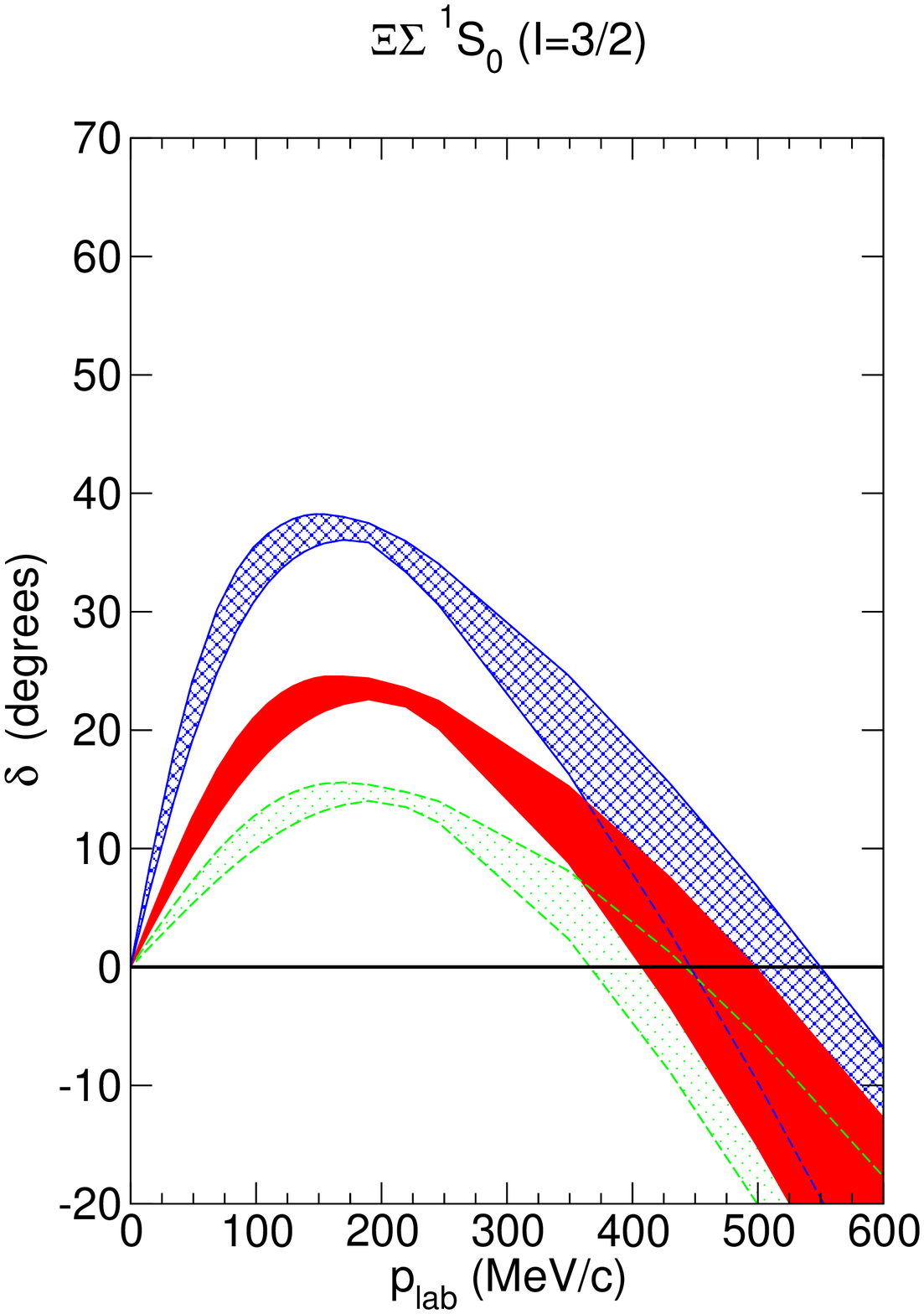}\includegraphics[height=8.0cm,keepaspectratio]{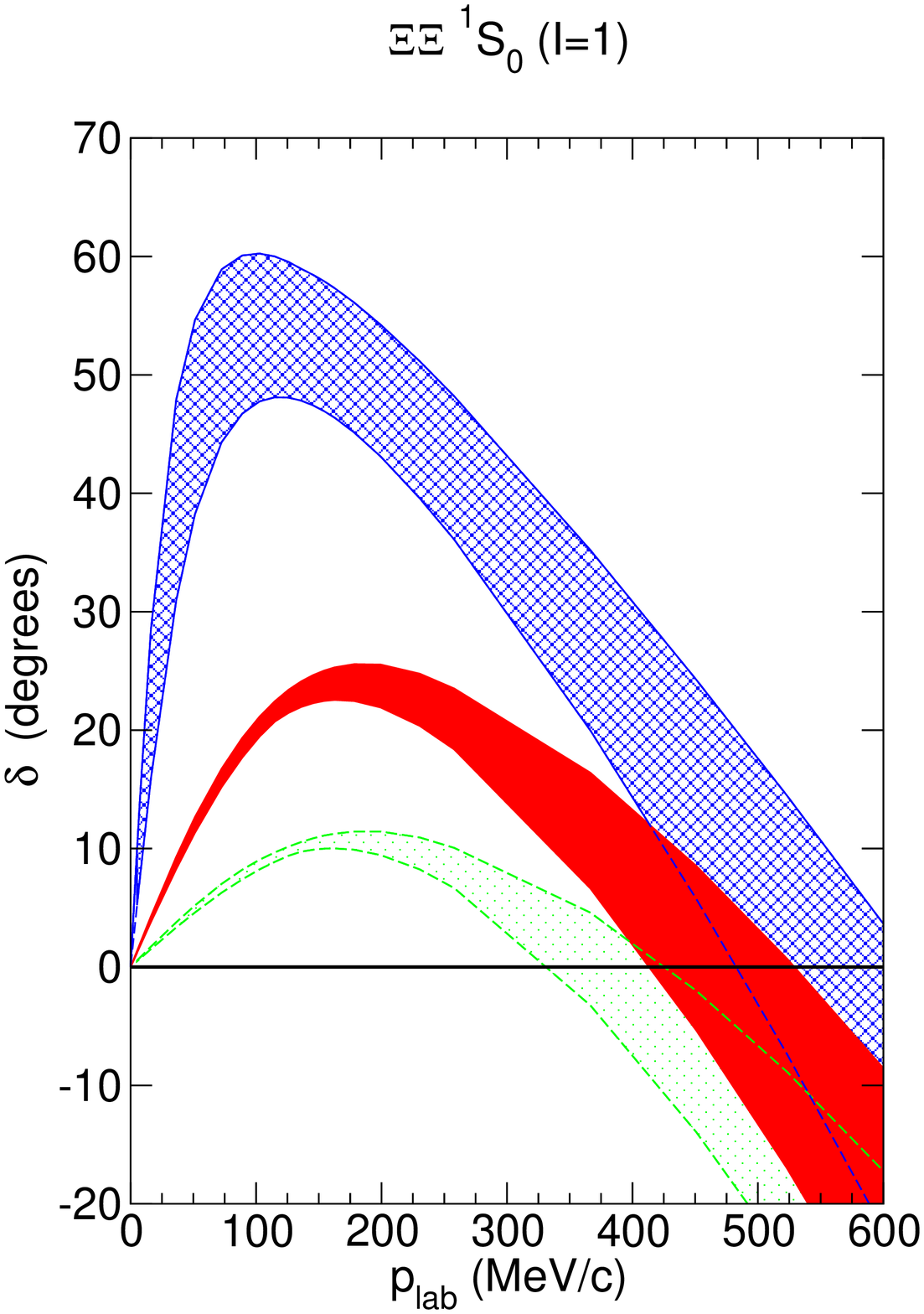}
\end{center}
\caption{Phase shifts for the $^1S_0$ partial wave in the $\Xi\Sigma$ ($I=3/2$) and $\Xi\Xi$ ($I=1$) channels. 
The hatched bands, filled bands, and dotted bands correspond to the choices $C^\chi_2 = 0$, $-C^\chi_1/2$, and
$-C^\chi_1$, respectively. 
}
\label{fig:XX}
\end{figure*}

The actual value of $C^\chi_2$ can be only determined by a fit to pertinent 
($\Xi Y$ and/or $\Xi\Xi$) data. Since such data are not available, in the 
following let us consider some exemplary choices for $C^\chi_2$.
In particular, we presume that the magnitude of the SU(3) breaking LEC $C^\chi_2$ 
is comparable to $C^\chi_1$ and that the trend in the SU(3) 
symmetry breaking we see for $NN \Rightarrow \Sigma N \Rightarrow \Sigma \Sigma$, 
is not reversed for the $S= -3$ and $-4$ systems. The latter means that we suppose 
$C^\chi_2$ to be positive, based on its definition via Eq.~(\ref{LEC}). 
A simple assumption is $C^\chi_2 \approx 0$, so that there is no further SU(3) 
symmetry breaking in the contact terms beyond $S= -2$.
The other extreme consists in assuming that $C^\chi_2 \approx -C^\chi_1$, which
implies that the same SU(3) symmetry breaking required to describe $pp$ and $\Sigma^+p$
occurs also between the $S= -2$, $-3$, and $-4$ $BB$ systems. 
Finally, we consider an intermediate case, namely $C^\chi_2 \approx -C^\chi_1/2$.
 
Predictions for the $\Xi \Sigma$ ($I=3/2$) and $\Xi\Xi$ ($I=1$) $^1S_0$ phase shifts 
resulting from the three choices are presented in Fig.~\ref{fig:XX}. 
In the case $C^\chi_2 \approx 0$ (hatched band) the only SU(3) symmetry breaking effects in 
the potential (as compared to $\Si\Si$) come from the one- and two-meson exchange contributions. 
One notices a clear increase in the attraction for $\Xi \Sigma$ and $\Xi\Xi$ in comparison 
to the $\Sigma\Sigma$ results, cf. Fig.~\ref{fig:SS} with Fig.~\ref{fig:XX}. Specifically, 
for $\Xi\Xi$ the phase shifts reach almost 60 degrees, i.e. similar values as in the
$pp$ system. 
Introducing an explicit SU(3) symmetry breaking in the contact terms
leads to the results represented by the filled bands ($C^\chi_2 \approx -C^\chi_1/2$) 
and dotted bands ($C^\chi_2 \approx -C^\chi_1$), respectively. Now the 
predicted phase shifts for the $^1S_0$ partial wave are much smaller and 
especially in the $\Xi\Xi$ case the reduction is drastic.

The scattering length for the $\Xi\Sigma$ channel are in the range of $-3.7$ to $-2.8$ fm 
for the choice $C^\chi_2 \approx 0$, but reduce to $-1.3$ to $-1.8$ fm 
for $C^\chi_2 \approx -C^\chi_1/2$, and to $-0.7$ to $-1.0$ fm for $C^\chi_2 \approx -C^\chi_1$.
For $\Xi\Xi$ we obtain $-7.0$ to $-13.5$ fm, $-1.6$ to $-1.8$ fm, and $\approx 0.7$ fm,
respectively. 

We have also performed calculations based on the $np$ $^1S_0$ phase shifts as starting point 
instead of the $pp$ values. In this case there is a somewhat stronger SU(3) symmetry breaking between 
$np$ and $\Sigma^+p$ and, accordingly, the resulting $\Si\Si$, $\Xi\Si$ and $\Xi\Xi$ phase shifts 
are then reduced by roughly 10 \% as compared to the ones presented in Figs.~\ref{fig:NN}, \ref{fig:SS} 
and \ref{fig:XX}. 
 
There are results for the $\Xi\Xi$ $^1S_0$ partial wave from lattice QCD calculations.  
The ones reported by the NPLQCD collaboration \cite{Beane} suggest a bound state with 
$E_B$ = -14.0 $\pm$ 1.4 $\pm$ 6.7 MeV. The calculation was performed for a pion mass 
of $m_\pi =$ 389 MeV and for $M_\Xi =$ 1349.6 MeV. 
In contrast, no bound state was found by the HAL QCD collaboration \cite{Sasaki}. In this
calculation, that corresponds to $m_\pi =$ 510 MeV and $M_\Xi =$ 1456 MeV, the 
interaction in the $^1S_0$ partial wave is only moderately attractive and the 
phase shifts rise only to a maximum of around 20$\pm$10 degrees. Interestingly, the 
EFT predictions based on the choice $C^\chi_2 \approx -C^\chi_1/2$ are fairly close to 
those results. 
Our investigations in Refs.~\cite{Haidenbauer11a,Haidenbauer11b} suggest that 
the actual value of the pion mass does not play an important role in the $\Xi\Xi$ system 
and, therefore, we do not expect sizable changes in the lattice results once 
calculations for masses closer to the physical value become feasible. The $\Xi$ mass is 
only marginally larger than the physical mass (which is about 1320 MeV) in case of the NPLQCD 
collaboration so that it should not distort the results. In any case, a smaller baryon 
mass would rather lead to a reduction of the attraction than to an enhancement,
cf. the discussion below. 

Finally, let us comment on the role played by the SU(3) symmetry breaking in the 
baryon masses. As mentioned in Sect. 2, the baryon masses enter solely in the course 
of solving the LS equation~(\ref{LSE}). Contributions to the potential involving
the baryon masses occur only at higher order in the employed power counting
scheme \cite{Wei90,Wei91,Epe05}. 
In fact, this appearance of the reduced $BB$ mass in the LS equation is the key point 
in the argument exploited by 
Miller \cite{Miller} in his exploration of possible $\Xi\Xi$ bound states. 
His argument is easy to understand in terms of the Schr\"odinger equation,
\begin{equation}
\nonumber
-\frac{d^2u}{d r^2} + 2 \mu_{B_1B_2} V_{B_1B_2}\, u = k^2\, u 
\end{equation}
where $u(r)$ is the wave function.
If SU(3) symmetry is approximately fulfilled then $V_{NN} \approx V_{\Xi\Xi}$.
However, since the physical mass of $\Xi$ is significantly larger than the one of 
the nucleon, the effective strength of the interaction is increased when it is multiplied 
with the appropriate reduced mass $\mu_{B_1B_2}$. For example, for $\Xi\Xi$ 
one has ${\mu_{\Xi\Xi}}/{\mu_{NN}} \approx 1.40$, i.e. there is a 40 \% increase 
in the effective strength of the interaction as compared to $NN$,
while for $\Sigma\Sigma$ one gets ${\mu_{\Sigma\Sigma}}/{\mu_{NN}} \approx 1.27$.
For attractive potentials this has a drastic effect and leads to bound states with 
increasing baryon masses as demonstrated in the work of Miller for simple potential models. 
Clearly, the same mechanism is also responsible for the $\Xi\Xi$, etc. bound states 
that one observes in meson-exchange potentials and in our EFT interactions 
when SU(3) symmetry is assumed in extrapolating to the strangeness $S= -3$ and $-4$ 
$BB$ systems. 
Indeed, the increase in the phase shift from $\Si\Si$ (Fig.~\ref{fig:SS}) to
$\Xi\Xi$ (Fig.~\ref{fig:XX}) in the scenario with $C^\chi_2 \approx 0$ reported
above is primarily dictated by the increase in the corresponding reduced masses.

The actual $pp$ and $\Sigma^+ p$ phase shifts suggest that there is no such net 
increase in the attraction when going to the strangeness sector. Thus, in practice 
the SU(3) symmetry breaking LEC $C^\chi_1$ (more than) compensates effectively the 
impact of the increase in the reduced mass. 
Indeed, the stepwise modification of the contact interaction due to the
SU(3) symmetry breaking terms that follows from chiral EFT, cf. Eq.~(\ref{LEC}), 
is paralleled by a similar stepwise increase in the reduced mass when going from 
$NN$ to $\Si N$ to $\Si\Si$, say. 
Thus, since the mass splitting between $\Sigma$ and $\Xi$ is significantly smaller  
than the one between nucleon and $\Sigma$, $M_\Xi - M_\Sigma \approx 125$ MeV 
versus $M_\Sigma - M_N \approx 254$ MeV, one could speculate that the magnitude of 
the ``compensating'' SU(3) symmetry breaking LEC ($C_2^\chi$) is likewise reduced. 
If that is so, adopting $C^\chi_2 \approx -C^\chi_1/2$ might be a reasonable choice.
In any case, we believe that a realistical estimation for $C^\chi_2$ might be 
provided by $-C^\chi_1/2 \ge C^\chi_2 \ge 0$.
But it is obvious from our results that for any value $C^\chi_2 \ge 0$ 
the bound states that we find in the $\Sigma\Sigma$, $\Xi\Sigma$ and $\Xi\Xi$ systems 
for interactions with SU(3) symmetric contact terms (cf. the results presented in
Sect. 3) disappear. 

\section{Summary}
In the present paper we examined the question whether baryon-baryon bound states in the 
strangeness sector could exist in the framework of chiral effective field theory. 
In particular, we explored the role of SU(3) symmetry breaking contact terms that arise 
at next-to-leading order in the perturbative expansion in the employed Weinberg scheme.
We focused on the $^1S_0$ partial wave and on baryon-baryon channels with maximal 
isospin because in this case there are only two independent SU(3) symmetry
breaking contact terms and, at the same time, those are the channels where
most of the bound states have been predicted in the past. 
Utilizing $pp$ phase shifts and $\Sigma^+ p$ cross section data allowed us to pin 
down one of the SU(3) symmetry breaking contact terms and a clear indication for the 
decrease of attraction when going from the $NN$ system to strangeness $S=-2$ 
is found, which practically rules out a bound state for the $\Si\Si$ $^1S_0$ partial
wave with isospin $I=2$. 
Furthermore, if that trend observed for $S=0$ to $S=-2$ is not reversed when going 
to the corresponding $\Xi\Sigma$ and $\Xi\Xi$ channels, which we assumed in the
present investigation, then also bound states in the latter systems are rather unlikely.

Experiments for $BB$ systems with $S=-3$ or $-4$ are certainly rather challenging.
However, it should be feasible to perform $\Xi\Si$ and $\Xi\Xi$ correlations measurements 
in heavy-ion collisions at RHIC or at CERN, similar to those for $\La\La$ reported 
recently \cite{Shah14}. From such data conclusions on the strength of the interaction 
in those systems could be drawn and possibly even on the existence of dibaryons. 
$BB$ systems with strangeness $S=-2$ to $-4$ could be also produced in photon induced 
reactions on the deuteron at JLab as suggested in Ref.~\cite{Miller}, or in 
corresponding $K^-$ induced reactions at J-PARC \cite{JPARC}. As discussed in 
\cite{Gasparyan12}, from such data one could even deduce the scattering lengths 
for specific $BB$ channels which would then provide a clear signal for the 
presence (or absence) of bound states. 
First and foremost, however, it would be good to resolve the discrepanices in the
present lattice QCD calculations for the $\Xi\Xi$ system. Hopefully, this can be 
done soon, because then we could get already a unique and definite answer.

\section*{Acknowledgements}

This work is supported in part by the DFG and the NSFC through
funds provided to the Sino-German CRC 110 ``Symmetries and
the Emergence of Structure in QCD'' and by the EU Integrated
Infrastructure Initiative HadronPhysics3. 

\appendix
\section{Two--pseudoscalar-meson exchange contributions}
\countzero

The spin-momentum part of the interaction in the $\Sigma\Sigma$, $\Xi\Sigma$ and 
$\Xi\Xi$ channels is the same as in the $YN$ case and is described in detail
in the Appendix A of Ref.~\cite{Hai13}. There are, however, some changes in the isospin
coefficents for $\Xi\Sigma$ as compared to $\Sigma N$ because the roles of $K$ and $\bar K$ 
and likewise of $N$ and $\Xi$ are interchanged. For convenience we summarize the isospin 
factors for the $\Xi\Sigma$ ($I=3/2$) case in Table~\ref{tab:isoXS}.
Those for $\Xi\Xi$ are identical to the ones for $NN$, with the 
replacement $N \leftrightarrow \Xi$ and $K \leftrightarrow \bar K$. 
The isospin factors for $\Si\Si$ and $I=2$ are given in Table~\ref{tab:isoSS}. 
Note that the isospin factors for the one-pseudoscalar-meson exchange can be found in 
Table 3 of Ref.~\cite{Hai10} (for $\Xi\Sigma$) 
and in Table 3 of Ref.~\cite{Po07} (for $\Si\Si$). 

The explicit SU(3) symmetry breaking in the decay constants is taken into account. 
The empirical values for these constants are \cite{PDG}
\begin{eqnarray}
\nonumber
f_\pi &=&  92.4 \ {\rm MeV}, \\ 
\nonumber
f_\eta&=&(1.19\pm 0.01) f_\pi, \\ 
f_K&=&(1.30\pm 0.05) f_\pi \ .
\end{eqnarray}
and we use the central values in our study. 
A somewhat smaller SU(3) symmetry breaking occurs also in the axial coupling constants,
see \cite{Ratcliffe,Yamanishi,Donoghue} but also \cite{Ber01,General}.
These effects are not taken into account in the present study. But 
we take the larger value $g_A = 1.29$ instead of $g_A = 1.26$ in order to
account for the Goldberger--Treiman discrepancy~\cite{Epe05}. 

As discussed in Appendix A.1 of Ref.~\cite{Hai13} the evaluation of the 
two-pseudoscalar-meson exchange gives also rise to a polynomial part. We 
assume here that those contributions only renormalize the LO and NLO contact terms 
and, therefore, they are not considered. Some of the terms omitted involve the 
masses of the pseudoscalar mesons and the SU(3) symmetry breaking generated by
them is assumed to be absorbed by the SU(3) symmetry breaking contact terms
$C^\chi_1$ and $C^\chi_2$.
In principle, there is also an SU(3) symmetry breaking due to differences 
in the baryon masses as discussed in Appendix B.2 of Ref.~\cite{Hai13}. 
However, since we consider here only channels with the same baryons in the
inital and final states, their effects are tiny and are not taken into 
account here. 

%
\begin{table*}
\caption{Isospin factors ${\mathcal I}$ for $\Xi\Si$ with $I=3/2$ for 
planar box, crossed box, triangle, and football diagrams consecutively. 
$B_{il}B_{ir}$ indicates the two baryons in the intermediate state and
$\pi\pi$ etc. the exchanged pair of mesons $M_1M_2$ for planar box and
crossed box diagrams. In case of the triangle diagrams there is only
a single baryon in the intermediate state.
See Ref.~\cite{Hai13} for details of notation. 
}
\begin{center}
\renewcommand{\arraystretch}{1.20}
\begin{tabular}[t]{|c||c|cccccc||c|ccc|}
\hline \hline
           & \MMBB \, & $\pi\pi$ & $\pi\eta$ & $\eta\pi$ & $\eta\eta$ & $ \pi K$ & $ \eta K$ 
& \MMBB\, & $\Kb \pi $ & $\Kb \eta $& $\Kb\,  \Kb $ \\
           & & & & & & & & & & & \\
\hline \hline
planar & $\Xi \Sigma $  & $1$ & $ 1$ & $ 1$ & $ 1 $ & $2$ & $ 2$ & $\Sigma\Xi$ & $ 2$ & $ 2$ & $4$ \\
box    & & & & & & & & & & & \\
\hline \hline
crossed   & $\Xi\Sigma $  & $3$ & $ 1$ & $ 1$ & $ 1 $ & $0$ & $0$ & $\Si\Si$ & $ 3$ & $ 2$ & $0$ \\
box       & $\Xi\Xi $     & $0$ & $0$ & $0$ & $0 $ & $2$ & $2$ & $\Si N$ & $ 0$ & $ 0$ & $2$ \\
          & $\Xi\Lambda $ & $2$ & $ 0$ & $0$ & $ 0 $ & $0$ & $ 0$ & $\La\La$ & $ 1$ & $ 0$ & $0$ \\
          &             & $ $ & $  $ & $  $ & $   $ & $ $ & $  $ & $\La N$ & $ 0$ & $ 0$ & $2$ \\
          &             & $ $ & $ $ & $  $ & $ $ & $ $ & $ $ & $\La\Si$,$\Si\La$ & $-1$ & $0$ & $0$ \\
\hline \hline
           & \MMB\, & $\pi\pi$ & $\pi K$ & $\eta K$ & $KK$ & &  
& \MMB \, & $\Kb \pi$ & $\Kb\eta$ & $\Kb\,\Kb$ \\ 
           & & & & & & & & & & & \\
\hline \hline
triangle  & $\Xi$  & $-8$ & $-2$ & $-2\sqrt{3}$&$0$ & & & $\Si$ & $ 4$ & $2\sqrt{3}$ & $ 2$ \\
right     &      &     &     &     &     & & & $\La$ & $-2$ & $0$ & $-2$ \\
\hline \hline
           & \MMB\, & $\pi\pi$ & $\pi \Kb$ & $\eta \Kb$ & $\Kb\,\Kb$ & &  
&  \MMB \,& $K \pi$ & $K\eta$ & $K K$ \\ 
           & & & & & & & & & & & \\
\hline
triangle  & $ \Si$ & $-2$ & $ 4$ & $2\sqrt{3}$ & $0$& & & $\Xi $ & $-2$ & $-2\sqrt{3}$ & $4$\\
left      & $\La $ & $-2$ & $-2$ & $0$ & $0$ & $$& & & $$& $$& $$\\
          & $N $ & $0$ &  $0$ & $0$ & $-8$& $$& & & $$& $$& $$\\
\hline \hline
football  & $ $    & $16$  & $ 6$ & $6$ & $ 4$ & & & $$ & $6$ & $6$  & $4$ \\
\hline \hline
\end{tabular}
\label{tab:isoXS}
\end{center}
\end{table*}

%
\begin{table*}
\caption{Isospin factors ${\mathcal I}$ for $\Si\Si$ with $I=2$ for 
planar box, crossed box, triangle, and football diagrams consecutively. 
$B_{il}B_{ir}$ indicates the two baryons in the intermediate state and
$\pi\pi$ etc. the exchanged pair of mesons $M_1M_2$ for planar box and
crossed box diagrams. In case of the triangle diagrams there is only
a single baryon in the intermediate state.
See Ref.~\cite{Hai13} for details of notation. 
}
\begin{center}
\renewcommand{\arraystretch}{1.20}
\begin{tabular}[t]{|c||c|cccc||c|cc|}
\hline \hline
           & \MMBB\, & $\pi\pi$ & $\pi\eta$ & $\eta\pi$ & $\eta\eta$ & \MMBB\, &
$K K$ & $\Kb\,\Kb$ \\
           &       & & & & & & &\\
\hline \hline
planar  & $\Si\Si$         & $ 1$ & $ 1$ & $1$ & $1$ & & & \\
box  & & & & & & & &\\
\hline \hline
crossed & $\Si\Si $ & $ 2$ & $ 1$ & $1$ & $1$ &$N N $ & $0$ & $4$ \\
box     & $\La\La $ & $ 1$ & $ 0$ & $0$ & $0$ &$\Xi\Xi $ & $ 4$ & $0$ \\
        & $\La\Si,\Si\La $ & $1$ & $0$ & $0$ & $0$ & & & \\
\hline \hline
triangle & $\Si $ & $-4$ & $0$ & $0$&$0$ & $N$ &  $0$ & $-4$  \\
left and right  & $\La$  & $-4$ & $0$ & $0$&$0$ & $\Xi$ & $-4$ & $0$   \\
\hline \hline
football & $    $  & $32$ & $0$ & $0$&$0$ & $ $ & $8$ & $8$  \\
\hline \hline
\end{tabular}
\label{tab:isoSS}
\end{center}
\end{table*}

\end{document}